\documentclass[a4paper,fleqn,usenatbib]{mnras}
\usepackage{newtxtext,newtxmath}

\usepackage[T1]{fontenc}
\usepackage{ae,aecompl}

\usepackage{graphicx}	
\usepackage{amsmath}	
\usepackage{amssymb}	
\usepackage[symbol]{footmisc}

\title[New water masers in IRAS 20231+3440]{Bowshocks in a newly discovered maser source in IRAS 20231+3440}
\author[C. S. Ogbodo et al.]{
Chikaedu S. Ogbodo,$^{1,3}$\thanks{E-mail: chikachukwuemeka@gmail.com}
R.A. Burns,$^{3,4,7}$
T. Handa,$^{3}$
T. Omodaka,$^{3}$
\newauthor
A. Nakagawa,$^{3}$
T. Nagayama,$^{2}$
M. Honma,$^{2}$
J. O. Chibueze$^{1,5,6}$
\newauthor
A. A. Ubachukwu,$^{1}$
and R. N. C. Eze$^{1}$
\\
$^{1}$Department of Physics and Astronomy, Faculty of Physical Sciences, University of Nigeria, Carver Building, 1 University Road, Nsukka, Nigeria\\      
$^{2}$Mizusawa VLBI Observatory, National Astronomical Observatory of Japan, 2-12 Hoshi-ga-oka, Mizusawa-ku, Oshu, Iwate 023-0861, Japan\\
$^{3}$Graduate School of Science and Engineering, Kagoshima University, 1-21-35 Korimoto, Kagoshima, Kagoshima 890-0065, Japan\\
$^{4}$Joint Institute for VLBI ERIC, Postbus 2, 7990 AA Dwingeloo, The Netherlands\\
$^{5}$SKA Africa, 3rd Floor, The Park, Park Road, Pinelands, Cape Town, 7405, South Africa\\ 
$^{6}$Space Research Unit, Physics Department, NorthWest University, Private Bag X6001, Potchefstroom, 2520, South Africa\\
$^{7}$Department of Physics/Geology/Geophysics, Federal University Ndufu-Alike, Ikwo, Ebonyi State, Nigeria
}

\date{Accepted XXX. Received YYY; in original form ZZZ}

\pubyear{2015}

\begin{document}
\label{firstpage}
\pagerange{\pageref{firstpage}--\pageref{lastpage}}
\maketitle


\begin{abstract}
From measuring the annual parallax of water masers over one and a half years with VERA, we present the trigonometric parallax and corresponding distance of another newly identified water maser source in the region of IRAS 20231+3440 as $\pi=0.611\pm0.022$ mas and $D=1.64\pm0.06$ kpc respectively. We measured the absolute proper motions of all the newly detected maser spots (30 spots) and presented two pictures describing the possible spatial distribution of the water maser as the morphology marks out an arc of masers whose average proper motion velocity in the jet direction was 14.26 km s$^{-1}$.
As revealed by the ALLWISE composite image, and by applying the colour-colour method of YSO identification and classification on photometric archived data, we identified the driving source of the north maser group to be a class I, young stellar object. To further probe the nature of the progenitor, we used the momentum rate maximum value (1.2$\times$10$^{-4}$ M$_{\odot}$ yr$^{-1}$ km s$^{-1}$) of the outflow to satisfy that the progenitor under investigation is a low mass young stellar object concurrently forming alongside an intermediate-mass YSO $\sim 60,000$ au ($\sim 37$ arcsecs) away from it.

\end{abstract}




\section{Introduction}
Early on, while still deeply embedded in their obscuring natal envelopes, young stellar objects (YSOs) at early evolutionary stages are difficult to investigate directly. However, even the youngest, most hidden YSOs experience protostellar ejections of mass driven by bi-polar jets which physically interact with ambient materials in their immediate environment. Such interactions produce signature tracers, making them easier to observe than the protostars driving them. Established correlations of certain outflow properties (see, for example, \citealt{Richer00, Beuther02}), make it possible to use observed properties of YSO ejections to investigate star formation mechanisms and even infer the nature of the associated YSO. In this work we use water maser observations of IRAS 20231+3440 as a diagnostic tool to infer the nature of a newly detected maser source in this region. \par

Water masers are associated with the early phases of both low and high mass YSOs, and because they are usually observed in shock regions where expelled materials collide with interstellar clouds, thus water masers are good outflow markers \citep{Marvel08}.
Very Long Baseline Interferometers (VLBI) have been extensively used to study masers becuase of their high resolving capabilities. Extended monitored observations of water masers by VLBI, yield accurate positions, line of sight velocities and skyplane movement of water masers. Water masers can therefore be used to reveal the 3D kinematics of shocked gas created by YSO outflows, making them an ideal for measuring outflow properties in young, obscured YSOs. \par

At the scales probed by VLBI within star-forming regions (a few au at a few kpc), little information on outflow morphology (e.g. width, length) is available, since the powering YSO and the rest of the environment are not usually observed with sufficient resolution nor astrometric accuracy. Water masers have been identified with the jet-disc interaction in a low mass YSO \citep{Moscadelli06} and with a rotating disc around an intermediate mass YSO \citep{Trinidad09}. In more cases, the masers clearly trace a bow shock \citep{Claussen98,Furuya00,Sanna12,Trinidad13,Burns15a,Burns16} and it can be argued that quantifiable physical properties like outflow velocity and outflow momentum rate can be most reliably derived from maser bowshocks. \par

IRAS 20231+3440, which is the region of study in this work, is an IRAS-identified young intermediate mass star-forming region with at least two previously identified young stellar cores SMM1 and SMM2, deeply embedded in the dust clouds of Lynds 870 \citep{Jiang04}. Water maser emission has been observed in IRAS 20231+3440 since 1993 \citep{Palla93}. Regarding tracers of star formation activity, broad line wing emission observed in the J=2--1 and 3--2 transitions of CO, dust continuum with $v_\mathrm{LSR}=+6.3$ km s$^{-1}$ \citep{Mao02} and radio continuum \citep{Xu13,Moscadelli16} showed that the SMM1 core has a bipolar outflow associated with it. \citet{Xu13} made radio VLBI astrometric observations of the water masers closely associated with the intermediate- mass YSO associated with IRAS 20231+3440 star forming region and presented $v_\mathrm{LSR}=+13.4$ km s$^{-1}$ and a parallax distance of $1.59\pm0.04$ kpc. \par

In Section 2 of this work we describe our recent VLBI water maser observations of IRAS 20231+3440, in which we find a new water maser source to the North of the intermediate mass YSO reported by \citet{Xu13}. In Section 3 we report that the new source is at a distance consistent with the intermediate mass YSO, as derived by annual parallax, and we show that the new source drives a water masers bowshock to the North-East direction. In Section 4 we use the 3D kinematics of the maser bowshock to derive the outflow properties of the new source, which are in turn used as a diagnostic tool to investigate the nature of the YSO. We confirm our diagnostics by comparison with a colour-colour analysis using archival infrared data. Finally, conclusions are given in Section 5.

\section{Observations and Data reduction}
Very Long Baseline Interferometry (VLBI) observations of IRAS 20231+3440, were carried out from the 26th of February 2013 to the 26th of September 2014 using the Japanese VLBI Exploration of Radio Astrometry (VERA). With VERA's unique installed dual-beam receivers \citep{Kawaguchi00}, it was possible to simultaneously observe the maser and continuum sources thus compensating the terrestrial effect in the VLBI phase referencing procedure \citep{Honma10} and eliminating the necessity for fast-switching. Atmospheric phase error due to water vapor in the troposphere is usually the main source of error that affects astrometry in VERA's K-band (22 GHz) \citep{Asaki07}, this error however, is much suppressed in VERA because it can observe two sources separated by up to 2.2 degrees simultaneously \citep{Kobayashi07}. The delay introduced by the hard-ware differences in the dual beam system is corrected by the injection of a characterized noise-signal \citep{Kawaguchi00}. For instrumental uncertainty, VERA can measure delay differences up to an accuracy of $\pm 0.1$ mm which is adequate for a positional accuracy of 10 micro-arcsec \citep{Honma08}.\par
Water masers in IRAS 20231+3440 and the continuum reference source, J2025+3343, were observed simultaneously without any switching. After initial data reduction of the first epoch data in which the position of the maser was unknown, we (re-)correlated all data at the position of the maser with the brightest flux as the phase tracking centre, which was set at
($\alpha,\ \delta)_\mathrm{J2000}=(20^\mathrm{h} 25^\mathrm{m} 07^\mathrm{s}.8013,\ +34^\circ 50^\prime 34^{\prime \prime}.7330)$, although \citep{Xu13} set it as ($\alpha,\ \delta)_\mathrm{J2000}=(20^\mathrm{h} 25^\mathrm{m} 07^\mathrm{s}.1053,\ +34^\circ 49^\prime 57^{\prime \prime}.539$), which is approximately 37 arcsec away from ours. \par
Seven epochs were conducted over the 18 month observation period. Each epoch lasted roughly 8 hours, and between the simultaneous observations of IRAS 20231+3440 and the phase reference source, the antenna beam was switched to a bright calibrator every 80 minutes for seven minutes. BL Lac, CTA102, DA55 and 3C454.3 were the standard calibrators used throughout the observations. \par 
Left-handed circular polarization signal was sampled using  2-bit  quantization,  and  filtered with the  VERA  digital filter unit \citep{Iguchi05}. With respect to the dual beam system, we term the beam centered on the maser target (at a rest frequency of 22.235 GHz) as the `A-beam', while the beam used for the continuum source is termed the `B-beam'.\\     
The bright continuum reference source, J2025+3343, in the B-beam, had 15 Intermediate Frequency (IF) channels assigned to it. Each channel had a bandwidth of 16 MHz and total frequency coverage was 240 MHz. One IF channel of 16 MHz was assigned to the line emission of target source, IRAS 20231+3440, in the A beam with 15.625 kHz frequency resolution, which corresponds to velocity spacing of $0.21$ km s$^{-1}$. \par
Correlation of the interferometric data was processed with the Mitaka FX correlator domiciled at the National Astronomical Observatory of Japan (NAOJ) Mitaka campus \citep{Chikada91}, using an accumulation period of 1 second. GPS measurements for atmospheric water vapor delays, applied to improve the delay tracking model in the post-correlation corrections \citep{Honma08}. \par 
Correlated data were reduced using the NRAO Astronomical Image Processing System (AIPS). Phase and rate solutions were obtained at 2 minutes intervals to solve atmospheric fluctuations along the line of sight to the reference source. We then used the direct phase reference method, with which calibrated solutions of the continuum reference source were applied to the maser sources. To identify the maser spots, we used the signal-to-noise-ratio (SNR) criteria, where we set the SNR of the peak intensity to be over 7 for all epochs. The date and resultant synthesized beam for each epoch is summarized in Table ~\ref{obs}.

\begin{table}
\begin{center}
\caption{Specification of our observations\label{obs}}
\begin{tabular}{lrrrrrr}
\hline
\multicolumn{1}{c}{Epoch ID}        &
\multicolumn{1}{c}{Date}   &
\multicolumn{1}{c}{Modified}      &
\multicolumn{1}{c}{Beam size} \\
 &
\multicolumn{1}{c}{DD/MM/YYYY} &
\multicolumn{1}{c}{Julian Date} &
\multicolumn{1}{c}{(mas$\times$mas, deg)} \\
\hline
1 & 26/02/2013 & 56349 & $1.20\times0.83, -50$ \\ 
2 & 26/04/2013 & 56408 & $1.28\times0.82, -44$ \\ 
3 & 04/06/2013 & 56447 & $1.34\times0.76, -58$ \\ 
4 & 02/11/2013 & 56598 & $1.26\times0.74, -52$ \\ 
5 & 09/02/2014 & 56697 & $1.24\times0.78, -48$ \\ 
6 & 09/05/2014 & 56786 & $1.27\times0.81, -50$ \\ 
7 & 26/09/2014 & 56926 & $1.28\times0.77, -52$ \\ 
\hline
\end{tabular}
\end{center}
\end{table}

\section{Results}
\subsection{Water maser detection and distribution in the star-forming region}
\begin{figure}

	\includegraphics[width=\columnwidth]{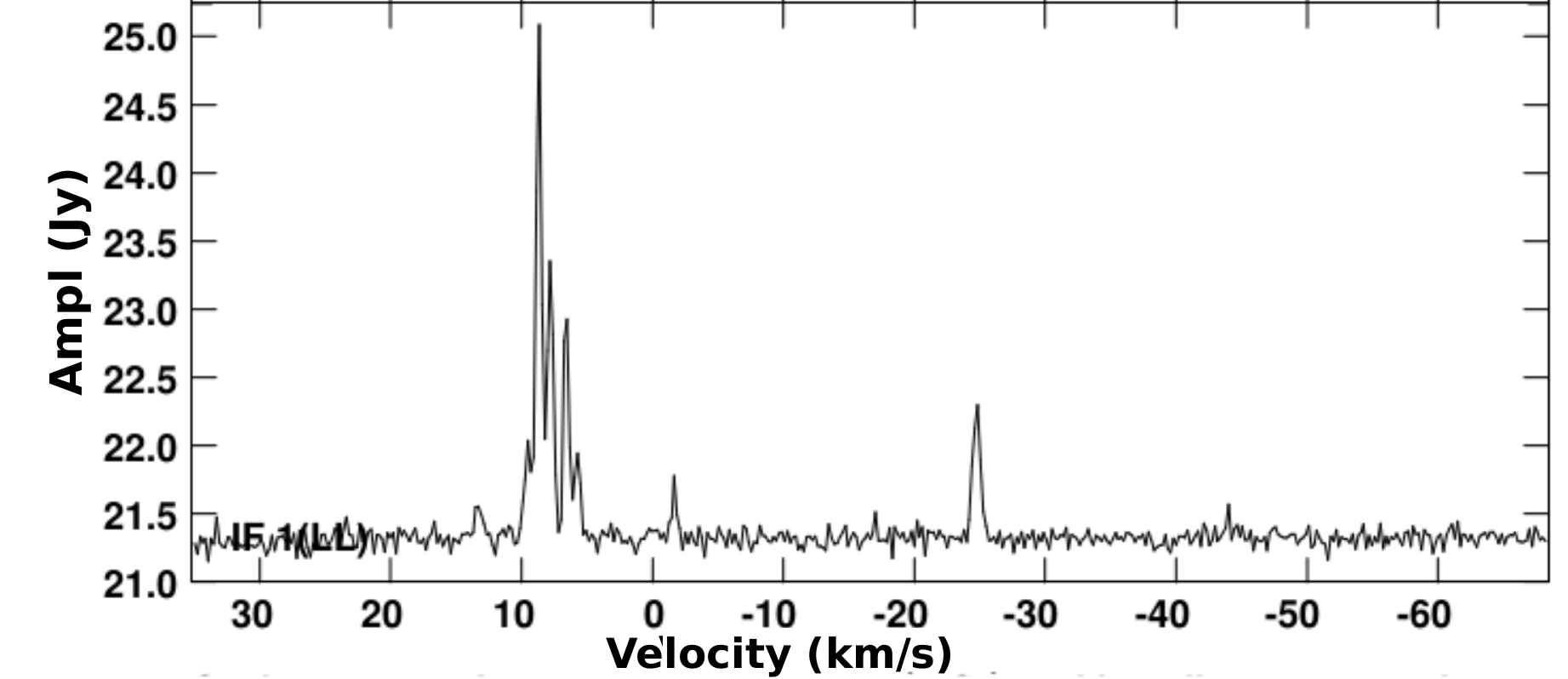}
    \caption{The scalar averaged cross-power spectrum as a function of time of maser emission for several averaged baselines in IRAS 20231+3440.}
    \label{spectrum}
\end{figure}

As seen from the intensity--velocity scalar averaged spectrum shown in Figure ~\ref{spectrum}, it is clear that there are several emission peaks. Maser emission was detected within an LSR velocity range of $+15$ km s$^{-1}$ to $-30$ km s$^{-1}$. \par
\begin{figure}
	\includegraphics[width=\columnwidth]{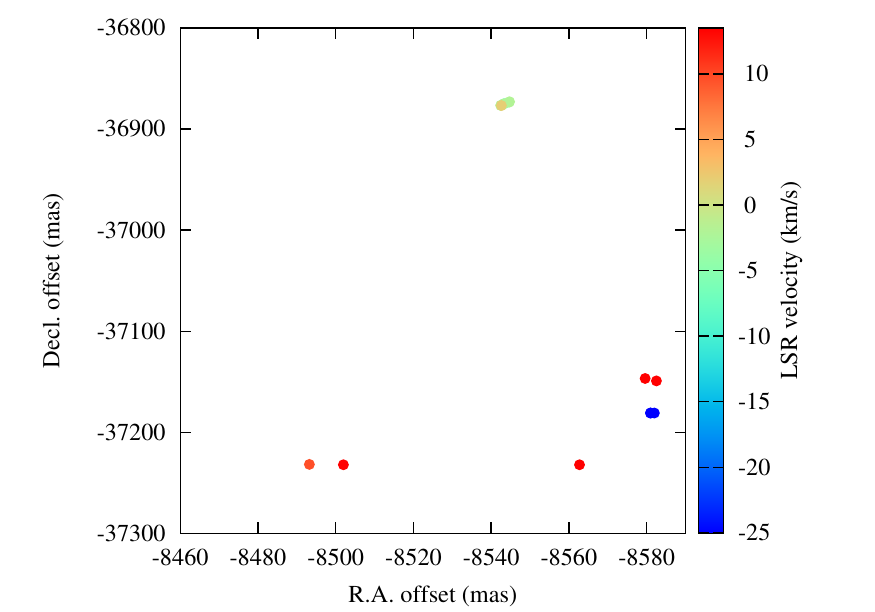}
    \caption{The water maser map of the south maser group. The map origin is $\alpha_{J2000.0}=20^{h}25^{m}07^{s}.8013$; $\delta_{J2000.0}=34{\degr}50{\arcmin}34{\arcsec}.7330$.}
    \label{southmasers}
\end{figure}

We identified two major maser groups, henceforth referred to as the North group and the South group. In many cases a gas clump has two or more maser spots. To identify these clumps, we define the maser ``feature" as maser spots with consistent radial velocity and occupy a position range not exceeding $1\times1$ mas of each other \citep{Burns14a}. We detected 30 maser spots in the North group which were grouped into 9 maser features based on spatio--kinematic proximity. We gave the position, proper motion, and radial velocity of each feature shown in Table ~\ref{secondtable}, as the equal weighted average of those of maser spots in the feature. Besides them, we also identified 8 maser spots in the South group shown in Table ~\ref{thirdtable}. Note that the South group corresponds to the same group of masers investigated by \citet{Xu13} while the North group are new detections. These two groups are well separated in the line-of-sight velocity domain. Spots in the North group show between 4 and 8 km s$^{-1}$ and correspond to the red-shifted complex emissions in Figure ~\ref{spectrum}. The other spots in the South group show about -25 and -2 km s$^{-1}$ in $v_\mathrm{LSR}$, which correspond to two blue-shifted peaks in Figure ~\ref{spectrum}.
We should note that the observed parameters of the maser spots in the south group shown in Figure ~\ref{southmasers} may be less accurate due to smearing effects caused by huge position offset from the phase tracking centre and long integration time. The separation between these two groups is about 37 arcsec, which corresponds to $\sim 60,000$ au, using the distance estimated from our parallax (see Section 3.2). At an offset of 37 arcsec, the reduction in sensitivity is slight and no correction was made for the primary beam. We think these two groups are driven by different energy sources (see Section 4.1.1) and we will focus on the North group only in this paper. Our only reference to the South maser group will be in Table ~\ref{thirdtable} and Figure ~\ref{southmasers} which report the positions and LSR velocities. \par

\subsection{Measurement of the parallax-distance}
\begin{figure}
\begin{center}
\includegraphics[scale=1.0]{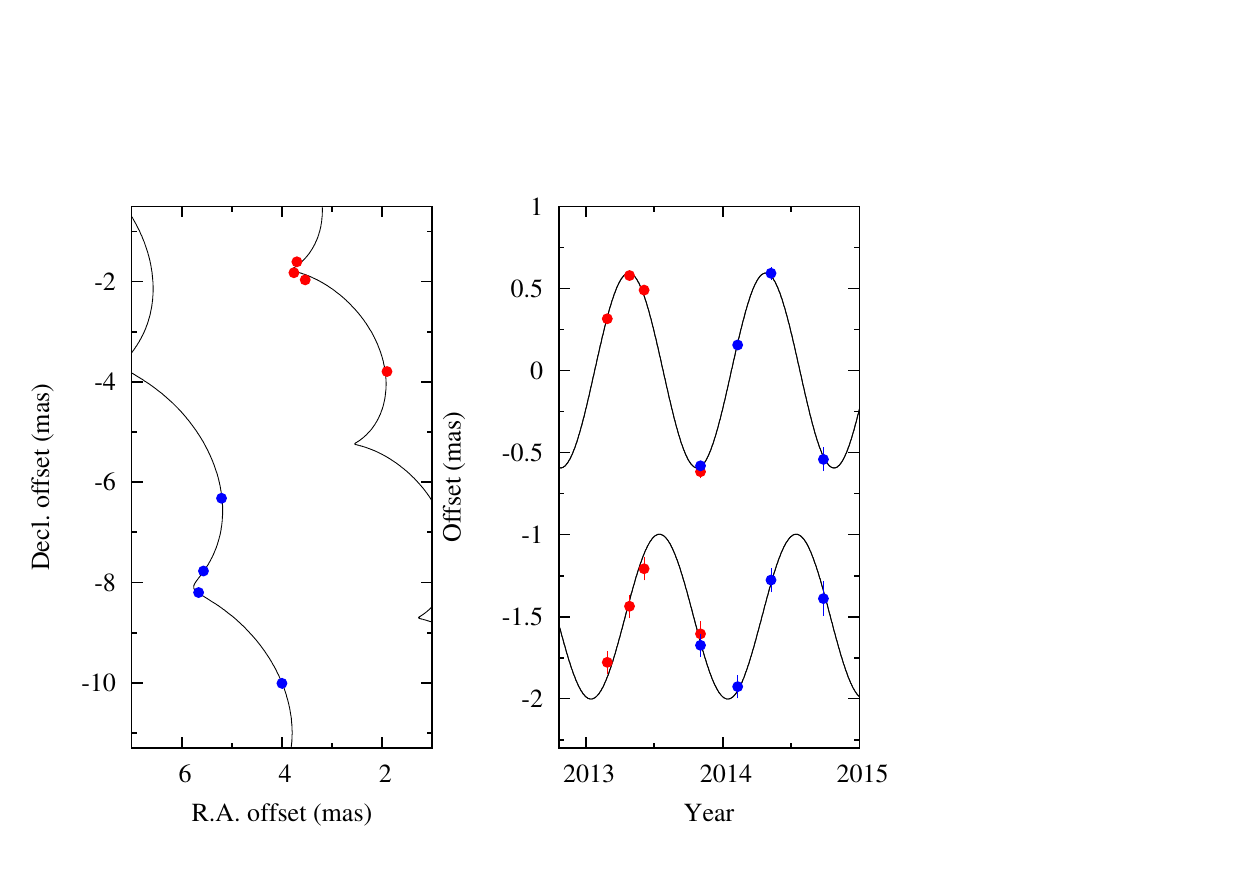} 
\end{center}
\caption{Parallax of IRAS 20231+3440. Red and blue spots represent spot No. 26 and 29, respectively. The left panel shows the motions of two spots on the sky. The right panel shows parallax motions of two spots along RA (upper) and Dec (lower) after subtracting linear motions on the sky. The linear motion of each spot was fitted independently, although the parallax motion of two spots was assumed to be the same.}
\label{parallax}
\end{figure}

The spots numbered 26 and 29 in Table ~\ref{secondtable} were detected at 4 epochs each, compared with 3 or fewer for all others, so we used these to fit for the parallax of the North group.
We assume these two spots are located at the same distance to reduce the error, since both spots are well fitted by a single parallax motion shown in Figure ~\ref{parallax}. We added and adjusted error floors to the fitting procedure to achieve a chi-square value of unity, these were 0.03 and 0.08 mas in RA and Dec., respectively. The chi-square fitting errors are consistent with the estimated noise-based position errors of the maser spots. \par 
We obtained a parallax of $0.611\pm0.022$ mas and a corresponding distance of $D=1.64\pm0.06$ kpc (uncertainty of 4 per cent). Given the offset between the two maser groups, the parallax distance obtained in this work, for the North group, is consistent with that obtained by those measured previously for the South group as  $D=1.59\pm0.05$ kpc \citep{Moscadelli16,Xu13}. 

\subsection{Proper and internal motions}
\subsubsection{Proper motions}
After removing the effects of parallax, the motions of the individual masers are affected by several factors: the bulk proper motion of the IRAS 20231+3440 system, the relative proper motions of the North and South clumps, the motions of each maser feature relative to the clump and the individual (e.g. turbulent/thermal) motion of each spot.\\
Absolute proper motions of masers were measured to fit thirty spots which were detected over three epochs. The parallax was fixed at 0.611 mas in this fitting. The resultant proper motions are listed in Table ~\ref{secondtable}. \par
Consistencies in spatial distribution and parallax distance between the two different driving sources and their associated maser cluster suggest that they are in the same star forming region. Without a symmetric distribution or randomness associated with the North maser group, it becomes almost impossible to evaluate the systemic proper motion from observed individual proper motions of the maser spots. By supposing that the relative motion between the two driving sources is small, it is assumed that the entire star forming region has a uniform group motion. Hence, we adopt the systemic proper motion of the South group for which \citet{Xu13} reported a proper motion of $(\mu_\alpha \cos \delta,\ \mu_\delta)=(-3.79\pm0.18, -4.88\pm0.25)$ mas yr$^{-1}$. 

\subsubsection{Internal motions}

\begin{figure}
\begin{center} 
\includegraphics[scale=1.0]{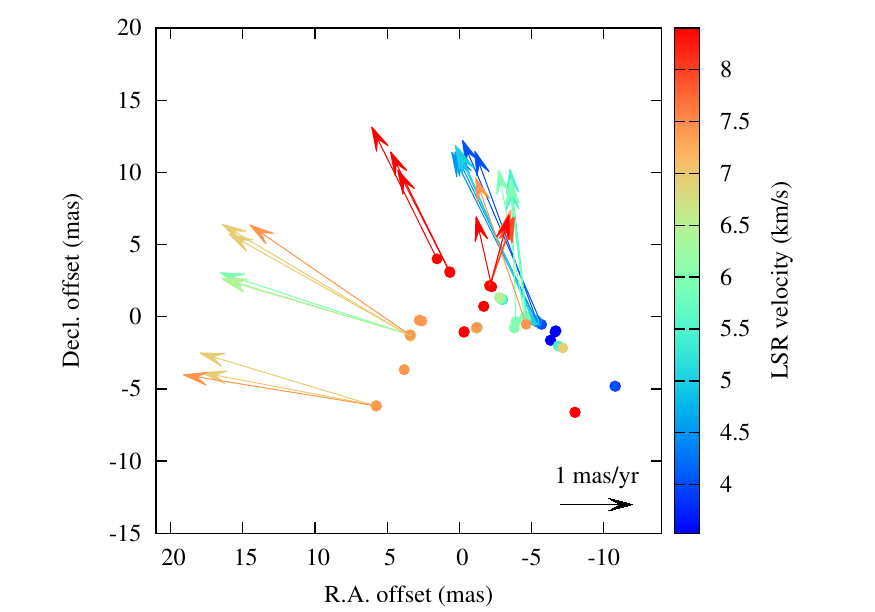} 
\end{center}
\caption{Internal motion of the north group after subtracting the systemic proper motion (Xu et al. 2013) of IRAS 20231+3440 region.}
\label{internalmotions_our}
\end{figure}

Maser internal proper motions, i.e. the motions of maser spots in the source frame, can be obtained by subtracting the systemic motion of the region from the measured absolute proper motions. Therefore, we converted the proper motions from Table ~\ref{secondtable} to the rest frame defined by the systemic motion from \citet{Xu13} (see Section 3.3.1), to give us the internal vector motions of the masers as graphically represented in Figure~\ref{internalmotions_our}. \par
\begin{figure}
	\includegraphics[width=7cm,height=18cm]{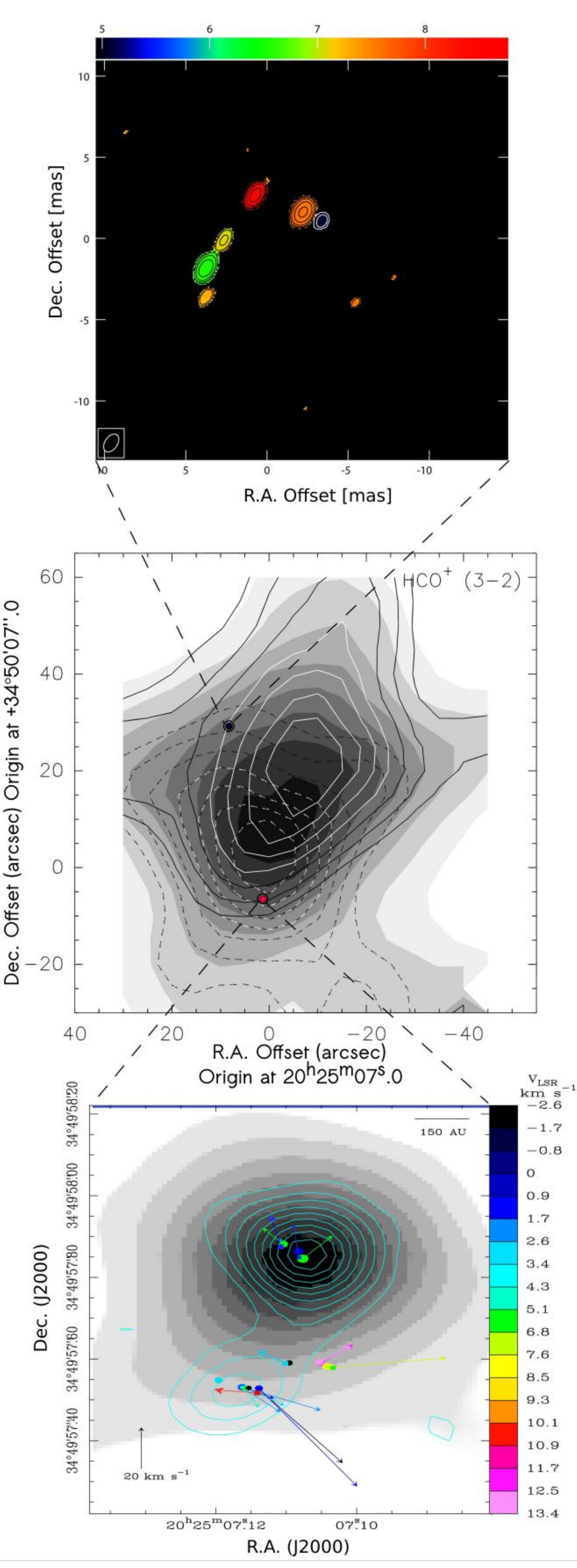}
    \caption{Overlays of maser positions. Top (North clump): Maser map of J2025+3450.  RA and Dec. at (0, 0) are $(20^\mathrm{h} 25^\mathrm{m} 07^\mathrm{s}.8013,\ +34^\circ 50^\prime 34^{\prime \prime}.7330)$. The colour bar represent the velocities of the maser features in km s$^{-1}$.} Middle: HCO$^+$ outflow map \citep{Mao02}. RA and Dec. at (0, 0) are $(20^\mathrm{h} 25^\mathrm{m} 07^\mathrm{s}.0,\ +34^\circ 50^\prime 05^{\prime \prime}.0)$ the black and red dots mark the positions of the North and South maser clumps. Bottom (South clump): JVLA C-band continuum of IRAS20231+3440\citep{Moscadelli16}.
    \label{overlays}
\end{figure}

 
The distribution and motion vectors in Figure ~\ref{internalmotions_our} (see also Figure ~\ref{overlays}), appear to trace a bow-shock propagating to the North-East and seem to be moving away from the unseen driving source. The jet position angle (PA) can be determined from the avarage PAs of all the maser motion vectors, we estimated this to be 27 degrees measured anti-clockwise from North. This agrees with the morphology of the bowshock as seen in the maser map. 
14.26 km s$^{-1}$ is the average of the component of proper motions in the jet direction of the bow shock masers. The average v$_{LSR}$ of the North clump is 6.92 km s$^{-1}$. \citet{Mao02} estimate a systemic v$_{LSR}$ of $\sim 6$ km s$^{-1}$ corresponding to a very small outflow velocity in the line of sight direction. Comparing the proper motion and line of sight velocities reveal that the bowshock propagates predominantly in the skyplane. This view conforms to the expected geometry possible of producing a bowshock appearance since masers amplify on surfaces tangential to the observer.


\begin{table*}
\begin{center}
\caption{Proper motion list of J2025+3450. The flux recorded is the maser's highest flux in all of the epochs it was observed. $(20^\mathrm{h} 25^\mathrm{m} 07^\mathrm{s}.8013,\ +34^\circ 50^\prime 34^{\prime \prime}.7330)$ is the RA and Dec. of (0,0). (AVR-Average, *-the spots used for parallax)}
\label{secondtable}
\begin{tabular}{rrrrrrrrc}
\hline
\multicolumn{1}{c}{Feature}                          &
\multicolumn{1}{c}{No}                          &
\multicolumn{1}{c}{$\Delta_\alpha$}             &
\multicolumn{1}{c}{$\Delta_\delta$}             &
\multicolumn{1}{c}{$\mu_{\alpha} \cos \delta$}  &
\multicolumn{1}{c}{$\mu_{\delta}$}              &
\multicolumn{1}{c}{$v_{\rm LSR}$}               &
\multicolumn{1}{c}{Flux}                    &
\multicolumn{1}{c}{Detection}                   \\

\multicolumn{1}{c}{ID}           &
                                         &
\multicolumn{1}{c}{(mas)}           &
\multicolumn{1}{c}{(mas)}           &
\multicolumn{1}{c}{(mas yr$^{-1}$)} &
\multicolumn{1}{c}{(mas yr$^{-1}$)} &
\multicolumn{1}{c}{(km s$^{-1}$)}   &
\multicolumn{1}{c}{(Jy/Beam)}       &
                                         \\
\hline
 1& 1 &$   -5.69\pm0.06 $&$   -0.54\pm0.13 $&$ -2.69\pm0.22 $&$ -2.33\pm0.41 $&$    4.18 $&$ 3.66\pm0.13 $&0001110 \\
  & 2 &$   -5.40\pm0.05 $&$   -0.36\pm0.13 $&$ -2.92\pm0.19 $&$ -2.52\pm0.39 $&$    4.39 $&$ 4.88\pm0.15 $&0001110 \\
  & 3 &$   -5.36\pm0.05 $&$   -0.32\pm0.13 $&$ -2.66\pm0.16 $&$ -2.53\pm0.39 $&$    4.60 $&$ 6.28\pm0.15 $&0001110 \\
  & 4 &$   -5.31\pm0.05 $&$   -0.25\pm0.13 $&$ -2.62\pm0.15 $&$ -2.55\pm0.38 $&$    4.81 $&$ 7.09\pm0.13 $&0001110 \\
  & 5 &$   -5.18\pm0.05 $&$   -0.26\pm0.13 $&$ -2.70\pm0.15 $&$ -2.46\pm0.38 $&$    5.02 $&$ 10.90\pm0.16 $&0001110 \\
  & 6 &$   -5.06\pm0.06 $&$   -0.23\pm0.13 $&$ -2.76\pm0.17 $&$ -2.51\pm0.39 $&$    5.23 $&$ 14.50\pm0.18 $&0001110 \\
\hline
  &AVR &$   -5.33 $&$   -0.32 $&$ -2.73\pm0.08 $&$ -2.48\pm0.16 $&$    4.71 $&         \\
\hline
 2& 7 &$   -4.63\pm0.06 $&$   -0.52\pm0.13 $&$ -3.09\pm0.17 $&$ -2.85\pm0.32 $&$    7.34 $&$ 16.00\pm0.35 $&0000111 \\
\hline
  &AVR &$   -4.63 $&$   -0.52 $&$ -3.09\pm0.17 $&$ -2.85\pm0.32 $&$    7.34 $&         \\
\hline
 3& 8 &$   -4.68\pm0.08 $&$   -0.09\pm0.14 $&$ -3.55\pm0.22 $&$ -2.82\pm0.40 $&$    5.44 $&$ 18.10\pm0.19 $&0001110 \\
  & 9 &$   -4.64\pm0.06 $&$    0.02\pm0.13 $&$ -3.56\pm0.16 $&$ -3.07\pm0.39 $&$    5.65 $&$ 14.10\pm0.17 $&0001110 \\
  &10 &$   -4.59\pm0.06 $&$    0.01\pm0.13 $&$ -3.58\pm0.17 $&$ -2.98\pm0.39 $&$    5.86 $&$ 6.79\pm0.13 $&0001110 \\
  &11 &$   -4.53\pm0.06 $&$   -0.01\pm0.13 $&$ -3.43\pm0.23 $&$ -2.84\pm0.41 $&$    6.07 $&$ 5.44\pm0.17 $&0001110 \\
\hline
  &AVR &$   -4.61 $&$   -0.02 $&$ -3.53\pm0.10 $&$ -2.93\pm0.20 $&$    5.75 $&         \\
\hline
 4&12 &$   -3.91\pm0.05 $&$   -0.34\pm0.12 $&$ -3.73\pm0.13 $&$ -2.91\pm0.31 $&$    6.07 $&$ 3.70\pm0.17 $&0000111 \\
\hline
  &AVR &$   -3.91 $&$   -0.34 $&$ -3.73\pm0.13 $&$ -2.91\pm0.31 $&$    6.07 $&         \\
\hline
 5&13 &$   -2.10\pm0.07 $&$    2.18\pm0.14 $&$ -4.08\pm0.37 $&$ -3.84\pm0.77 $&$    7.13 $&$ 11.10\pm0.32 $&1110000 \\
  &14 &$   -2.07\pm0.05 $&$    2.18\pm0.13 $&$ -4.13\pm0.29 $&$ -3.93\pm0.73 $&$    7.34 $&$ 16.00\pm0.35 $&1110000 \\
  &15 &$   -2.08\pm0.05 $&$    2.16\pm0.13 $&$ -4.14\pm0.28 $&$ -3.94\pm0.72 $&$    7.55 $&$ 16.10\pm0.34 $&1110000 \\
  &16 &$   -2.10\pm0.05 $&$    2.14\pm0.13 $&$ -4.11\pm0.28 $&$ -3.93\pm0.73 $&$    7.76 $&$ 12.70\pm0.52 $&1110000 \\
  &17 &$   -2.13\pm0.05 $&$    2.12\pm0.13 $&$ -4.05\pm0.30 $&$ -3.90\pm0.73 $&$    7.97 $&$ 8.46\pm0.21 $&1110000 \\
  &18 &$   -2.24\pm0.08 $&$    2.07\pm0.15 $&$ -3.57\pm0.43 $&$ -3.91\pm0.79 $&$    8.18 $&$ 7.39\pm0.12 $&1110000 \\
\hline
  &AVR &$   -2.12 $&$    2.14 $&$ -4.01\pm0.15 $&$ -3.91\pm0.30 $&$    7.66 $&         \\
\hline
 6&19 &$    0.68\pm0.08 $&$    3.12\pm0.16 $&$ -3.07\pm0.40 $&$ -3.47\pm0.83 $&$    7.97 $&$ 8.45\pm0.12 $&1110000 \\
  &20 &$    0.65\pm0.51 $&$    3.06\pm0.13 $&$ -2.96\pm0.28 $&$ -3.21\pm0.73 $&$    8.18 $&$ 2.89\pm0.19 $&1110000 \\
\hline
  &AVR &$    0.66 $&$    3.09 $&$ -3.02\pm0.24 $&$ -3.34\pm0.56 $&$    8.07 $&         \\
\hline
 7&21 &$    1.55\pm0.14 $&$    4.01\pm0.34 $&$ -2.89\pm0.30 $&$ -3.05\pm0.73 $&$    8.39 $&$ 7.39\pm0.12 $&1110000 \\
\hline
  &AVR &$    1.55 $&$    4.01 $&$ -2.89\pm0.30 $&$ -3.05\pm0.73 $&$    8.39 $&         \\
\hline
 8&22 &$    3.35\pm0.06 $&$   -1.28\pm0.14 $&$ -1.15\pm0.34 $&$ -4.01\pm0.76 $&$    6.07 $&$ 5.30\pm0.16 $&1110000 \\
  &23 &$    3.37\pm0.05 $&$   -1.26\pm0.13 $&$ -1.18\pm0.29 $&$ -4.11\pm0.73 $&$    6.29 $&$ 7.71\pm0.12 $&1110000 \\
  &24 &$    3.38\pm0.05 $&$   -1.27\pm0.13 $&$ -1.18\pm0.28 $&$ -4.09\pm0.72 $&$    6.50 $&$ 45.90\pm0.52 $&1110000 \\
  &25 &$    3.37\pm0.04 $&$   -1.35\pm0.11 $&$ -1.18\pm0.16 $&$ -3.33\pm0.34 $&$    6.71 $&$ 50.40\pm0.58 $&1111000 \\
  &26*&$    3.39\pm0.04 $&$   -1.32\pm0.09 $&$ -1.27\pm0.12 $&$ -3.47\pm0.26 $&$    6.92 $&$ 51.00\pm0.86 $&1111000 \\
  &27 &$    3.43\pm0.07 $&$   -1.26\pm0.13 $&$ -1.57\pm0.14 $&$ -3.36\pm0.30 $&$    7.13 $&$ 8.99\pm0.32 $&1110000 \\
\hline
  &AVR &$    3.37 $&$   -1.30 $&$ -1.26\pm0.09 $&$ -3.73\pm0.21 $&$    6.60 $&         \\
\hline
 9&28 &$    5.79\pm0.06 $&$   -6.19\pm0.14 $&$ -1.36\pm0.20 $&$ -4.14\pm0.40 $&$    6.71 $&$ 4.71\pm0.24 $&0001110 \\
  &29*&$    5.79\pm0.05 $&$   -6.15\pm0.10 $&$ -1.39\pm0.12 $&$ -4.43\pm0.23 $&$    6.92 $&$ 29.80\pm0.53 $&0001111 \\
  &30 &$    5.74\pm0.05 $&$   -6.17\pm0.13 $&$ -1.11\pm0.19 $&$ -4.45\pm0.41 $&$    7.13 $&$ 7.99\pm0.27 $&0001110 \\
\hline
  &AVR &$    5.77\ $&$   -6.17 $&$ -1.29\pm0.12 $&$ -4.34\pm0.20 $&$    6.92 $&         \\
\hline
\end{tabular}
\end{center}
\end{table*}

\begin{table}
\begin{center}
\caption{Positions and V$_{LSR}$ velocities of the South masers.}
\label{thirdtable}
\begin{tabular}{rrrrc}
\hline
\multicolumn{1}{c}{No}                          &
\multicolumn{1}{c}{$\Delta_\alpha$} &
\multicolumn{1}{c}{$\Delta_\delta$}             &
\multicolumn{1}{c}{$v_{\rm LSR}$}               &
\multicolumn{1}{c}{Flux}                        \\
&
\multicolumn{1}{c}{(mas)}           &
\multicolumn{1}{c}{(mas)}           &
\multicolumn{1}{c}{(km s$^{-1}$)}   &
\multicolumn{1}{c}{(Jy/Beam)}       \\
\hline
  1 &$   -8542.46 $&$   -36876.62 $&$    -1.67 $ & $ 2.65\pm0.13 $\\
  2 &$   -8542.55 $&$   -36876.63 $&$    -2.09 $ & $ 1.01\pm0.09 $ \\
  3 &$   -8542.35 $&$   -36876.87 $&$    -0.82 $ & $ 7.36\pm0.41 $ \\
  4 &$   -8580.96 $&$   -37181.04 $&$    -23.9 $ & $ 0.88\pm0.09 $ \\
  5 &$   -8580.98 $&$   -37181.04 $&$    -24.2 $ & $ 1.84\pm0.11 $ \\
  6 &$   -8581.01 $&$   -37181.02 $&$    -25.1 $ & $ 3.40\pm0.15 $ \\
  7 &$   -8543.32 $&$   -36874.94 $&$    -4.41 $ & $ 6.63\pm0.38 $ \\
  8 &$   -8544.72 $&$   -36873.05 $&$    -3.35 $ & $ 9.80\pm0.61 $ \\

\hline
\end{tabular}
\end{center}
\end{table}

\begin{figure}
	\includegraphics[width=\columnwidth]{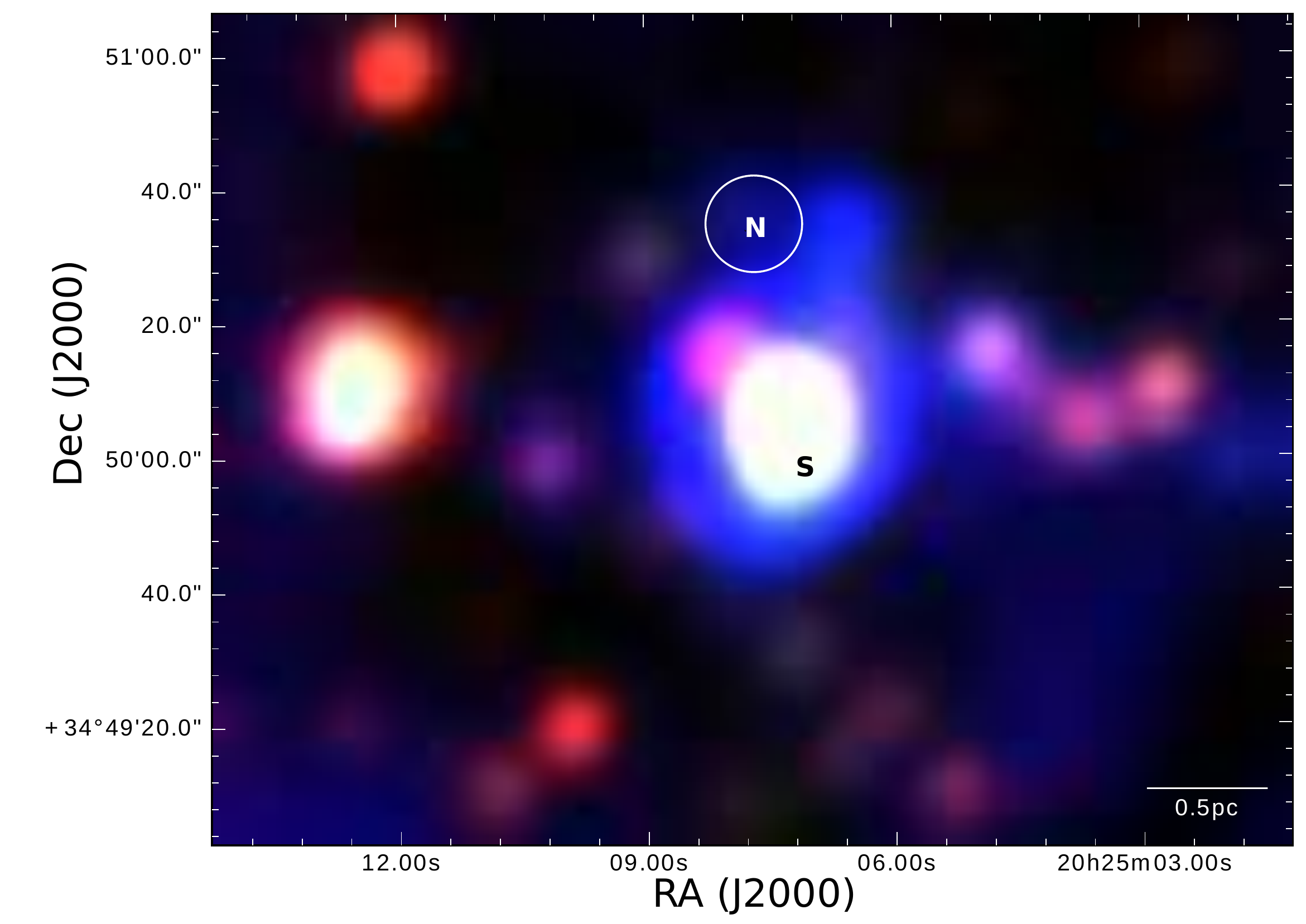}
    \caption{Composite image of the {\it WISE} bands (blue: 3.4 $\micron$, blue: 4.6 $\micron$, green: 12 $\micron$, red: 22$\micron$) showing a faint infrared source, J2025+3450 (circled), driving the north masers. "N" roughly marks out the position of the North group masers, and "S" also showing the position of south group within the bright object (intermediate-mass IRAS20231+3440) just below the circle.}
    \label{allwise}
\end{figure}

\section{Discussion}
\subsection{Star formation activity in IRAS 20231+3440}
Previous publications on the IRAS 20231+3440 region concentrate on the intermediate-mass YSO which dominates the emission in this region \citep{Moscadelli16, Xu13}. The aforementioned intermediate-mass YSO has associated masers \citep{Xu13} which are seen in the VERA data, and are referred to as the South group in this paper. The angular separation of the centroids of the North and South maser groups is $37\pm0.05$ arcsec. Assuming that both groups are at about the same distance along the line of sight, the North group is at least 60000 au away from the well-studied intermediate-mass YSO; a separation too large to be reconciled by scenarios involving only a single driving source. Thus, the newly detected masers reported in this paper associate with another YSO in the same star forming region. In the following sections we discuss the previously unexplored North source based on the VERA maser data and archival photometric data.

\subsubsection{Identifying the driving source}
Searching the Vizier database several infrared sources in the ALLWISE catalogue were found in the IRAS 20231+3440 region. Figure ~\ref{allwise} shows a composite-band image from combining all FITS file of all the {\it WISE} wavebands. The emission in this region is evidently dominated by the intermediate mass YSO (brightest source below the circle in Figure ~\ref{allwise}), which is surrounded by several less intense infrared objects. One {\it WISE} infrared object, J202507.78+345034.5 (hereafter J2025+3450), encircled in Figure ~\ref{allwise}, is spatially consistent with the North maser group, having co-ordinates $\alpha_{J2000.0}=20^{h}25^{m}07^{s}.789$; $\delta_{J2000.0}=34{\degr}50{\arcmin}34{\arcsec}.55$; an offset of 0.24 arcseconds from the maser phase tracking center.
The ALLWISE mid-infrared magnitudes of J2025+3450 are 15.3, 12.4, 9.0, and 2.8 mag at the 3.4, 4.6, 12, and 22 $\micron$ bands, respectively. Based on the photometry of J2025+3450, colour-colour analysis for of this object \citep{Koenig14} corresponds to a class I YSO. The colour-colour analysis is able to distinguish the emission between YSOs and other foreground/background, enabling us to confirm the protostellar nature of the infrared source. \par

Due to the uncertainty in the {\it WISE} astrometric accuracy (0.237 arcseconds) we can only constrain the relative positions of the masers and infrared source to within 0.44 arcsecond of each other; the exact determination of the separation and position angle between the North masers and the ALLWISE source is not possible to any higher accuracy. A maximum separation of up to 440 mas corresponds to a physical separation of up to 720 au or less; i.e. a few hundred au. This is a typical order of separation seen between the shock fronts of young YSO jets and their driving sources \citep{Goddi11, Torrelles11}, and we could thus expect that the arc shape of the masers was formed in a bowshock tracing the front of a protostellar jet  \citep{Furuya00, Burns16}. On the other hand, in the case that the North group masers are much closer to the ALLWISE object then the maser morphology would be better explained as an expanding ring \citep{Moscadelli06, Trinidad09} for low-mass YSO and \citep{James16, Torrelles11} for high-mass YSOs, with the driving source located near the center of the maser map; (0,0). In either case it appears reasonably certain that the North group masers are produced in shocks related to ejection or expansion motions emanating from J2025+3450.

\subsubsection{Investigating the driving source with maser kinematics}
22 GHz water masers are excited under warm (few hundred - $\sim 1200$ K), fairly dense ($\sim 10^{10}$ -- $\sim 10^{8}$ cm$^{-3}$) conditions \citep{Neufeld91, Gray16}. Moreover, water molecules re-form rapidly after shock dissociation. Thus, in star-forming regions, 22-GHz masers are predominantly found in the shocks associated with ejections and expansion motions near young stars. In the case of a protostellar jet, the physical properties of the jet relates to the properties of the driving source. Thus, the 3D motions of masers can be used as a diagnostic to infer the properties of the YSO \citep{Goddi11}.

The momentum rate of an outflow is given by $\dot{P}=1.5\times10^{-3}V^{2}_{10}R^{2}_{100}(\frac{\Omega}{4\pi})n_{g}$ M$_{\odot}$ yr$^{-1}$ km s$^{-1}$ \citep{Goddi11}. Where $V_{10}$ is the maser velocity in units of 10 km s$^{-1}$, R$_{100}$ is the outflow length (given by the separation between the maser emission and the driving source) in units of 10 au, $\Omega$ is the jet opening angle, and n$_{8}$ is the ambient density in units of 10$^{8}$ cm$^{-3}$ for which we consider a range of 10$^{7}$ to 10$^{9}$ cm$^{-3}$ \citep{Hollenbach13}. As explained in Section 3.3.2, we take the velocity of the bow shock as 14.26 km s$^{-1}$, the average of the internal maser proper motions (assuming that the exciting source shares the systemic velocity of the South group).
Since we have no direct information on the length of the outflow, we trial a range of possible values, namely 1, 10, 100, 1000 and 10000 au. We derive the jet opening angle from the fixed jet width (24.6 au, based on the maser morphology) and the trial value of the jet length.
The parameters and resulting range of momentum rates for each tested outflow length are shown in Table ~\ref{table4}.

\begin{table}
\begin{center}
\caption{Momentum rate outcomes and associated opening angles for different jet-lengths.\label{table4}}
\begin{tabular}{lrrrr}
\hline
\multicolumn{1}{c}{S/N}        &
\multicolumn{1}{c}{Trialled}   &
\multicolumn{1}{c}{Jet opening}      &
\multicolumn{1}{c}{Momentum rate} \\
 &
\multicolumn{1}{c}{jet-length (AU)} &
\multicolumn{1}{c}{angle (sr)} &
\multicolumn{1}{c}{M$_{\odot}$ yr$^{-1}$ km s$^{-1}$} \\
\hline
1 & 1 & 1.74 & $4.2\times10^{-9}-4.2\times10^{-7}$ \\ 
2 & 10 & 1.07 & $2.6\times10^{-7}-2.6\times10^{-5}$ \\ 
3 & 100 & 0.05 & $1.2\times10^{-6}-1.2\times10^{-4}$ \\ 
4 & 1000 & 0.0005 & $1.2\times10^{-6}-1.2\times10^{-4}$ \\ 
5 & 10000 & 0.000005 & $1.2\times10^{-6}-1.2\times10^{-4}$ \\ 
\hline
\end{tabular}
\end{center}
\end{table}

Initially, the resulting jet momentum rate increases for longer trial values of the jet length, however at increasing jet lengths the corresponding jet opening angle narrows. As a result, the increasing momentum rates plateau by the balancing changes of R and $\Omega$. Therefore, for any feasible jet length the maximum value of the momentum rate is 1.2$\times10^{-4}$ M$_{\odot}$ yr$^{-1}$ km s$^{-1}$.
According to the outflow parameter correlations in \citet{Beuther02}, mechanical force (momentum rate) is correlated both to the bolometric luminosity and mass of the core. An approximate value of the progenitor luminosity can be estimated from the corresponding order of magnitude of the momentum rate. Therefore, by adopting the $\dot{P}$ - L$_{bol}$ and the $\dot{P}$-M$_{core}$ relationship of \citet{Beuther02}, we can see that even our highest momentum rate corresponds to a YSO of luminosity of 100 L$_{\odot}$, at most. 
This momentum rate calculation assumes that the relative motion between the north driving source and the source driving the south clump is approximately zero. While this assumption cannot be tested without observation of a counter-jet to confirm the systemic motion of the North source, a relative proper motion of 10 km s$^{-1}$ would produce a factor of $\sim 4$ difference i.e. less than one order of magnitude. \par

To investigate the alternative maser-progenitor configuration, which posits an expanding ring, we calculated the dynamical age of the probable expanding ring. This case assumes that the driving source is at the centre of surrounding masers marking out gas motion of the envelope. The driving source is stationary as the masers uniformly move away from it with a common radial expanding velocity of 16.55 km s$^{-1}$ (proper motion average of maser spots in all direction). At a distance of 1.64 kpc and an expanding ring of radius of 8.2 AU, the dynamical age of the expanding ring was calculated to be $\sim 2$ yr. The dynamical timescale of the proposed expanding ring observed by VERA is similar to those seen in some MYSOs in the literature e.g. $\sim 8$ yr for Cepheus A HW \citep{Torrelles11} and $\sim 12.5$ for M17 \citep{James16}. Therefore, based only on the spatio-kinematics of the observed masers alone an expanding ring structure is one possible conclusion. In this case it would represent the first known case of an expanding ring traced by masers in a low mass, class I YSO. \par


Stellar winds from massive young stellar objects (MYSOs) release copious amount of mechanical energy in the molecular clouds sufficient to drive masers in an expanding shell morphology \citep{Lada87}. Early phases of MYSOs are energetic enough to produce stellar winds which drive the expanding shell associated with them. Such early activity is possible within the framework of massive star formation since MYSOs ignite nuclear fusions well before dispersing the gas and dust coccoons in which they form. Conversely, in the case of low mass stars, nuclear fusion does not commence until the later stages of evolution (class III YSOs and T-Tauri onwards) \citep{Schulz12}, thus the class I low-mass YSO discussed in this work might be too young to produce the stellar winds required to drive an expanding shell. 

\subsubsection{Evolutionary stage of the driving source}
In the previous subsections we have shown that the North group masers associate with an infrared source, J2025+3450, which a colour-colour analysis reveals to be a class I YSO. Furthermore, our analysis of the jet momentum rate revealed that the driving source can be of mass no greater than 100 L$_{\odot}$, which is consistent with that obtained from the infrared photometry. In conclusion, the North group water masers are associated with a very young, low mass YSO which has yet to be investigated prior to this work.

\subsection{Simultaneous formation of low and high-mass stars}
Theoretical differences between low- and high-mass star formation and evidence of some similarities in physical observations of common evolutionary phases, have resulted to a persistent difficulty in clearly resolving the different formation pathways. High mass star formation which is less understood than the low-mass star formation has two major formation theories- the monolithic collapse model and competitive accretion model. The monolithic collapse/core accretion model is more like an upscaled version of low mass star formation as it describes the gravitational collapse of cores whose initial mass determines the eventual mass of the post-formation star. While the competitive accretion model decribes the formation of high mass stars as a result of favorable position within clusters \citep{Zinnecker07}. The competitve accretion model predicts the possibility of simultaneous formation of both low and high mass stars. Simulations predicted the possibilty of coincidental formation of massive stars and star clusters habouring a population of low mass young stellar objects \citep{Smith09}. An example is the high resolution, sub-millimeter observations of the simultaneous high-mass and low-mass  formation in G11.92-0.61 \citep{Cyganowski17}. Here, we interestingly point out the discovery of a low mass YSO in the vicinity of an intermediate-mass YSO. The SMM1 core of IRAS 20231+3440 is an intermediate-mass class I YSO \citep{Jiang04,Moscadelli16} while the concurrently forming J2025+3450 is a low-mass class I YSO.


\section{Conclusions}
Results from 18 months VERA observations of 22 GHz water masers in the star forming regions of IRAS 20231+3440 led us to detect a new group of 30 water maser spots associated with another driving-source. From which we also determined a trigonometric parallax of $\pi=0.611\pm0.022$ mas corresponding to a distance of $D=1.64\pm0.06$ kpc, consistent with the parallax distance reported by \citet{Xu13}. We also measured and presented the absolute proper motions and vector representation of all the maser internal motions. \par
Colour-colour photometric analysis of {\it WISE} data revealed a class I YSO as the north maser group progenitor. By estimating the momentum rate of the outflow, We used results from the maser bowshock diagonistics to confirm the outcome of the {\it WISE} data analysis. The maximum value of the momentum rate (1.2$\times$10$^{-4}$ M$_{\odot}$ yr$^{-1}$ km s$^{-1}$) correspond to that of a low-mass star luminosity. Thus, we concluded that the YSO is a low mass YSO. \par
Note that the bowshock interpretation is the most probable, based on known YSO properties, but the assumption that the driving source shares the South group systemic velocity is purely circumstantial.




\bibliographystyle{mnras}
\bibliography{Complete_manuscript_file} 








\bsp	
\label{lastpage}
\end{document}